\documentclass[doublecol]{epl2} 

\usepackage{amsmath}
\usepackage{color}
\usepackage{amsfonts}
\usepackage{amssymb}
\usepackage{breqn}
\usepackage{graphicx}
\usepackage{array}
\usepackage[utf8]{inputenc}
\usepackage{hyperref}
\usepackage{float}
\usepackage{xcolor}
\usepackage{subfig}
\usepackage{subcaption}
 \date{}

\title{A Brief Study of Dark Energy Accretion onto Schwarzschild Black Hole : Biswas-Roy-Biswas Type Redshift Parameterization is Chosen}
\shorttitle{A Brief Study of Dark Energy Accretion onto Schwarzschild Black Hole} 

\author{Subhajit Pal\inst{*}\footnote{subhajitpal968@gmail.com$~;~~\text{Orchid}~:~0009-0002-2431-1375$} \and Sukanya Dutta\inst{**}\footnote{sduttasukanya@gmail.com$~;~~\text{Orchid}~:~0009-0004-2989-309X$} \and Ritabrata Biswas\inst{**}\footnote{biswas.ritabrata@gmail.com$~;~~\text{Orchid}~:~0000-0003-3086-892X$}}
\shortauthor{Subhajit Pal \etal}

\institute{                    
  \inst{*} Department of Mathematics, Jadavpur University, Kolkata-32, India\\
  \inst{**} Department of Mathematics, The University of Burdwan, Burdwan-713104, India}


\abstract{
In this letter, we have considered accretion of a particular type of Dark Energy model onto a Schwarzschild type black hole. Before using the model, the free parameters of the Dark Energy model have been constrained with differential ages data. A narrow peak on top of a wide plateau in two parameters' distributions indicates a well defined best fit value embedded within a broad region of near-degenerate solutions. This means the data strongly favours one specific parameter value but also permit a wide range with comparable likelihood. Physically, it reflects that the Dark Energy dynamics are locally constrained yet globally insensitive to small parameter variations. An increasing $\log_{10}\left[M(z)/M_{0}\right]$ since $z=3$ signifies that black holes have continuously grown through accretion and mergers within the standard hierarchical formation scenario. The precise rate of this growth depends on the radiative efficiency $\epsilon$, the
effective accretion parameter $\lambda_{\rm eff}$, and the cumulative impact of merger events.}

\begin{document}

\maketitle


If we consider the cosmological constant $\Lambda$, dark energy (DE) turns to stay constant in space and time following $T_{\mu\nu}=-\rho_{\Lambda}g_{\mu\nu}$ \cite{faraoni2007cosmological}. Quintessence, on the contrary, varies slowly like a scalar field, and in principle, is able to accrete onto black holes (BHs). Due to a small gradient of scalar field, effective inflow of energy is negligible to keep the BH mass growth due to DE accretion practically undetectable \cite{jacobson1999primordial}. Babichev et. al. \cite{babichev2004dark} have shown, in phantom era, i.e., when $1+\omega<0$, accretion rate turns negative, i.e., $\dot{M}=4\pi AM^2(1+\omega)\rho<0$, $\omega=\frac{p}{\rho}$ is the equation of state(EoS) of DE. $M$ and $\dot{M}$ are the mass and rate of change in mass of BH respectively. Thus phantom energy accretion leads a BH to loose mass. In an extreme scenario, (``Big Rip" universe \cite{bouhmadi2005escaping}), BHs evaporate completely as the phantom energy density diverges. 

The mathematical necessity of redshift parameterization arises from the fact that the evolution of DE enters the Friedmann dynamics through the equation \cite{Chevallier2001}
\begin{equation*}
    \rho_{DE}(z) = \rho_{DE,0} \times \exp\left\lbrace{3 \int_0^z \frac{1 + \omega(z^\prime)}{1 + z^\prime}dz^\prime}\right\rbrace~~~,
\end{equation*}

Since \(\omega(z)\) is unknown and cannot yet be derived from a fundamental Lagrangian, one must prescribe a functional form for it. A redshift parameterization is enough perfect to make the integral analytically solvable and the model comparable with observational data. Expressing \(\omega\) as a function of \(z\) (for example, \(\omega(z) = \omega_0 + \omega_a \frac{z}{1 + z}\) \cite{Chevallier2001, Linder_2003})  allows the resulting \(\rho_{DE}(z)\) and Hubble rate \(H(z)\) to be written in closed form, enabling direct confrontation with measurable quantities such as luminosity distances and  Baryon acoustic oscillations (BAO) scales. Thus, parameterization is not merely an empirical convenience but a mathematical framework that translates the unknown time evolution of DE into an integrable and observable function of redshift.

In this letter, we recall an ansatz for the functional form of the energy density of DE, $\rho_{DE}$ as \cite{Biswas:2019ayj}
\begin{equation}
    \frac{1}{\rho_{DE}}\frac{d\rho_{DE}}{da}=-3\left[\frac{\lambda_1}{1+ak_1}+\frac{\lambda_2(1-a)}{(1+ak_2)^2}\right]~~~,
\end{equation}
where, $\lambda_1,~\lambda_2,~k_1~\text{and}~k_2$ are constants. 

Integrating, we get the following equation:
\begin{equation}\label{rho}
    \rho_{DE}=A_0\frac{(1+ak_2)^{\frac{3\lambda_2}{k_2^2}}}{(1+ak_1)^{\frac{3\lambda_1}{k_1}}}\exp{\left\{{\frac{3\lambda_2(1+k_2)}{k_2^2(1+ak_2)}}\right\}}~~~,
\end{equation}
where, $A_0=\rho_{DE0}\frac{(1+k_1)^{\frac{3\lambda_1}{k_1}}}{(1+k_2)^{\frac{3\lambda_2}{k_2^2}}} \exp{\left\lbrace{-\frac{3\lambda_2}{k_2^2}}\right\rbrace}$ and $\rho_{DE0}$ is the present time (at $z=0$) value of the scalar field density. Equation \eqref{rho} gives the energy density function of the Biswas-Roy-Biswas(BRB) DE model.

Using equation \eqref{rho}, we get the EoS parameter $\omega_{\phi}$\cite{Biswas:2019ayj} as a function of redshift ($z=\frac{1}{a}-1$) as
\begin{equation}\label{Omega}
    \omega_{\phi}(z)=-1+\frac{\lambda_1}{(1+k_1)+z}+\frac{\lambda_2 z}{\left\{(1+k_2)+z\right\}^2}~~~~.
\end{equation}
One of the existing popular redshift parametrization models, Chevallier-Polarski-Linder (CPL) generically produces a non-negligible early DE fraction unless priors are imposed on one of its free parameters, $w_a$. BRB suppresses dark energy dynamics automatically at early times, avoids tension with Big Bang Neucleosynthesys(BBN) and recombination constraints. CPL and another popular and renowned redshift parametrization model Jassal-Bagla-Padmanabhan model (JBP) are Taylor-like expansions in redshift variables. As a result, CPL can show rapid transitions near $z\sim 1$ and JBP forces an artificial peak in evolution around intermediate redshift. BRB model is non-polynomial, ensures smooth evolution over all $z$. This avoids sharp transitions that lack physical motivation. CPL often crosses $w=-1$ unintentionally, leading to theoretical instability in scalar-field realizations. JBP restricts crossing too strongly. BRB controlled phantom crossing (or its avoidance), a parameter-dependent behavior that can be mapped to stable effective theories. CPL/JBP parameters are purely phenomenological. Our EoS admits a clearer interpretation : $\lambda_1$ implements low-$z$ deviation from $\Lambda$ and $\lambda_2$ interprets higher-order dynamical correction.

For a homogeneous, isotropic universe, the Friedmann Lemaitre Robertson Walker (FLRW) metric is the following 
\begin{equation}
    ds^2 = -dt^2 + a(t) ^2 \left[{\frac{dr^2}{1-k r^2}+r^2 d \Omega^2 }\right]~~~,
\end{equation}
where, the parameter $k$ denotes the spatial curvature, which can take values $-1,~ 0,~ 1$ corresponding to an open, flat, or closed universe respectively. (The function $a(t)$ is known as the cosmic or Robertson–Walker scale factor.) $d\Omega^2=d\theta^2+sin^2\theta d\phi^2$ is the unit sphere. Assuming that dark matter(DM) behaves as a perfect fluid, the Friedmann equation can then be expressed in the following form as c=1 and cosmological contant $\Lambda=0$
\begin{equation}\label{Hubble parameter}
    H^{2} = \frac{8\pi G}{3}\rho -\frac{k}{a^2}~~~~.
\end{equation}
Here, $H$ is the Hubble parameter, and $\rho$ is the energy density in the usual meaning. We rewrite the equation \eqref{Hubble parameter} as,
\begin{equation}\label{Hubble_1}
    H^2 = \frac{8 \pi G}{3} \left(\rho_b + \rho_{DE}+ \rho_r\right)- \frac{k}{a^2}~~~,
\end{equation}
where $\rho_b$, $\rho_{DE}$ and $\rho_r$, respectively, are the density contributions from baryons, DE and radiation.

For the expansion rate, we can define
\begin{equation}
\begin{split}
    E^2(z) \equiv \frac{H^2 (z) }{H_0^2} = \Omega_{b0} (1+z)^3 &+\Omega_{DE0} F(z) + \Omega_{r0} (1+z)^4 \\
    &+ \Omega_{k 0}(1+z)^2~~~,
    \end{split}
\end{equation}
where $\Omega_{b0},~\Omega_{DE0}~\text{and}~\Omega_{r0}$ are the present day dimensionless densities which take the form
\begin{equation}
    \Omega_i \equiv \frac{8 \pi G}{3 H_0^2} \rho_i~~,~~~i=\text{rad, DE and DM}~~~.
\end{equation}
Curvature contributes an additional term of
\begin{equation}
    \Omega_k \equiv - \frac{k}{H_0^2 a^2}~~~~.
\end{equation}
$F(z)$ will be governed by the form of DE that has been chosen. 
As $E(z=0)=1$ at the present epoch,  we derive 
\begin{equation}
    \Omega_{b0} + \Omega_{DE0}+ \Omega_{r0}+\Omega_{k 0}=1~~~~.
\end{equation}
 Within the framework of this model, we shall construct a methodology for estimating the Hubble parameter across a redshift parameterization.

 Using the definition of the Hubble parameter at redshift $z$, i.e., $H(z)=\frac{\dot{a}}{a}$ and the relation between the redshift $z$ and the scale factor $a$, we differentiate with respect to cosmic time,
$\frac{dz}{dt}=\frac{d}{dt}(1+z)=-\frac{\dot{a}}{a^2}=-\frac{H(z)}{a}=-H(z)(1+z)\implies H(z)=-\frac{1}{1+z}\frac{dz}{dt}~~$.

Starting from the fundamental equation of the different ages method, we note that the exact derivative $\frac{dz}{dt}$ cannot be precisely measured  directly from observations. We approximate it using finite differences based on nearby data points which gives $\frac{dz}{dt} \approx \frac{\Delta z}{\Delta t}~~$.

Stellar population synthesis models applied to galaxy spectra are used to determine their ages. The analysis focuses on massive, passively evolving galaxies with minimal star formation, selected within a narrow redshift range. From these, class two galaxies are chosen at specific redshifts and age $\left(z,~t(z)\right)$ and $\left(z+\Delta z , t(z + \Delta z)\right)$ (say), respectively. Then $H(z) \approx -\frac{1}{1+z}\frac{(z+\Delta z)-z}{t(z+\Delta z)-t(z)}~~$, assuming $\Delta z = z_2 -z_1 \ll 1$.

As the Hubble parameter is estimated through a ratio, the dominant source of uncertainty arises from $\Delta t$.
\begin{align}
    H(z) = -\frac{1}{1+z} \frac{\Delta z}{\Delta t} \implies \delta H(z) = \frac{1}{1+z} \frac{\Delta z}{\Delta t^2} \delta(\Delta t) ~~~.
\end{align}

This approach enables a direct measurement of the Hubble parameter $H(z)$, in contrast to integral methods such as type Ia supernovae (SNeIa) and BAO, which provide constraints on cosmological distances. Importantly, it does not require prior assumptions about spatial flatness, the nature of DE, or specific cosmological models. Consequently, it offers a powerful means of distinguishing between different DE models through the redshift evolution (i.e., slope) of $H(z)$.

To quantify the uncertainty of measurement, we define the uncertainties in redshift and differential age as $\sigma_{\Delta z}$ and $\sigma_{\Delta t}$ respectively. These are then used to propagate the error in the estimation of $H(z)$ as $\sigma_H^2 = \left( \frac{\partial H}{\partial \Delta z} \right)^2 \sigma^2 _{\Delta z}+ \left( \frac{\partial H}{\partial \Delta t} \right)^2 \sigma^2 _{\Delta t}~~$.

Hence, the partial derivatives are derived as 
\begin{align}
    \frac{\partial H}{\partial \Delta z} = -\frac{1}{1+z} \frac{1}{\Delta z}~~~\text{and}~~~ \frac{\partial H}{\partial \Delta z} = \frac{1}{1+z} \frac{\Delta z}{\Delta t^2}~~~.
\end{align}

We leaving the total uncertainty expressed as,
$\sigma^2_H = \left(\frac{1}{1+z} \frac{1}{\Delta t} \right)^2 \sigma^2 _{\Delta z} + \left(\frac{1}{1+z} \frac{\Delta z}{\Delta t^2} \right)^2 \sigma^2 _{\Delta t}~~~.$

Using differential ages method, different Hubble parameter data are obtained which are listed in table 1 of \cite{Biswas:2025bve} and the references therein.
The chi-squared statistic quantifies the deviation between observed and theoretical expansion rates, with each squared difference normalized by the corresponding observational uncertainty for Observed Hubble Data (OHD),
\begin{equation}
\chi^2_{OHD}(H_0,~\tilde{\theta})= \sum_{i=1}^{N} \frac{\left\lbrace{H_{th} \left({z_i, \tilde{\theta}}\right)-H_{obs} (z_i)}\right\rbrace^2 }{\sigma_zi^2}~~~,
\end{equation}
where $H_{obs}(z_i)$ is the observed value of the Hubble parameter at redshift $(z_i)$ and  $H_{th} \left({z_i, \tilde{\theta}}\right)$ is the theoretical value from a model depending on parameters $\tilde{\theta}$ (e.g. $H_0$, $\Omega_m$, $\lambda_1$, $\lambda_2$ $k_1$ and $k_2$ etc.). $\sigma_{z,i}^2$ denotes the $1\sigma$ uncertainty  in the observed $H(z_i)$. $N$ is the total count of data points.

Also from the reference \cite{Eisenstein_2005}, we get the expression for $A(z)$ which is commonly observable and for $\chi^2_{BAO}$ respectively. Again using Planck 2018 compressed likelyhood \cite{Planck2018}, we obtain the expression for $R = 1.7492 \pm 0.0049$
CMB shift parameter and the WMAP$7$ data gives us $R=1.726 \pm 0.018$ at $z=1091.3$ \cite{Eisenstein_2005, Planck2018, Wang_2013, Shafer_2015, Jimenez_2002, Wang_2007}. Chi-squared is also defined, in this case, as 
\begin{equation}
    \chi^2 _{CMB} = \frac{(R-0.469)^2}{(0.017)^2}~~~.
\end{equation}
The total chi squared function for joint analysis turns out 
\begin{equation}
    \chi _{tot}^2 = \chi^2_{OHD}+\chi^2_{BAO}+ \chi^2_{CMB}~~~.
\end{equation}
Therefore, the best fit values of different DE model parameters are obtained from Table1, as we have calculated this particular values in this table.
\begin{figure}[!ht]
    \centering
    \subfloat[For H(z)-z data Analysis] {\includegraphics[height=1.7in,width=1.7in]{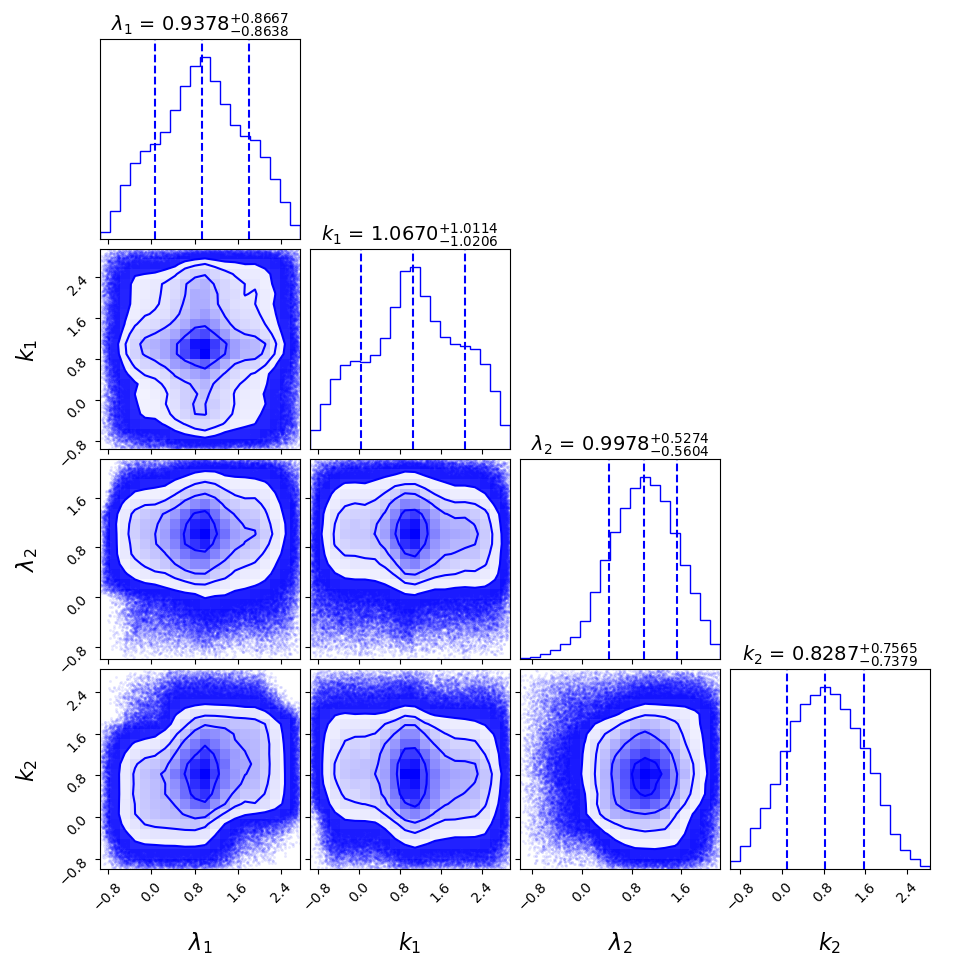}}
   \subfloat[For H(z)-z+BAO Analysis] {\includegraphics[height=1.7in,width=1.7in]{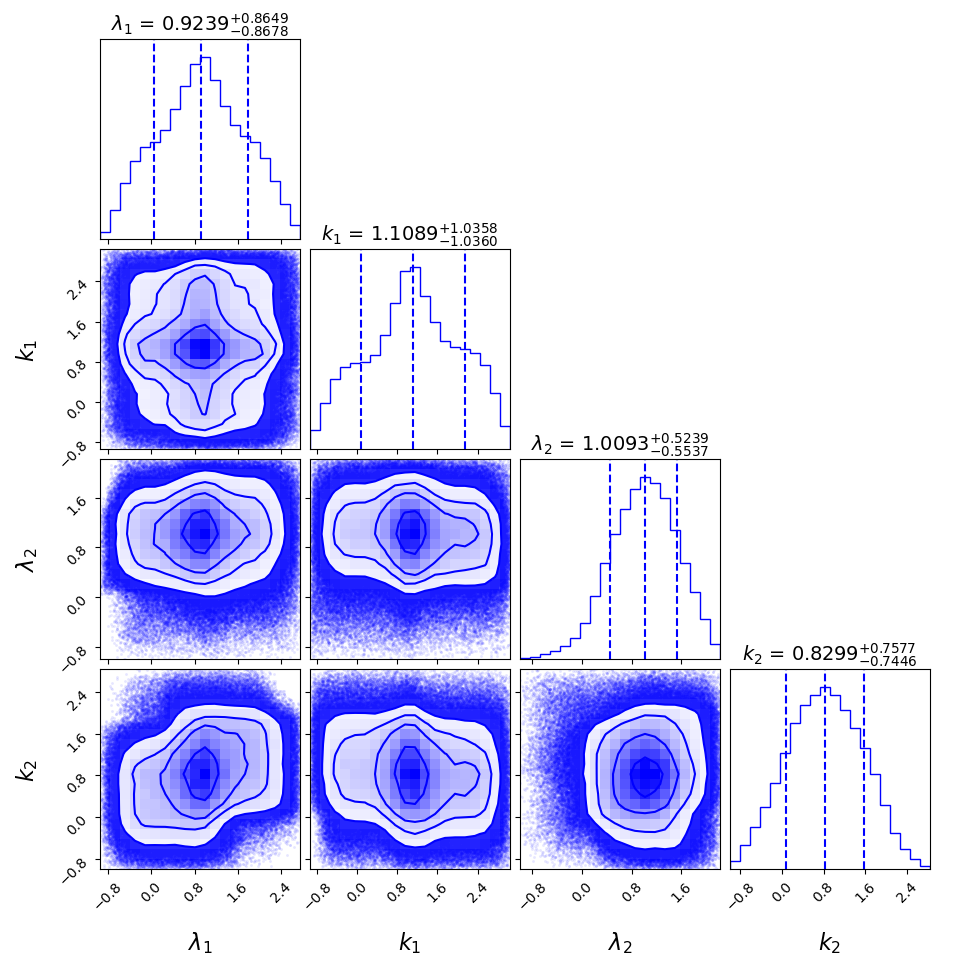}} \\
    \subfloat[For H(z)-z+BAO+CMB Analysis] {\includegraphics[height=1.7in,width=1.7in]{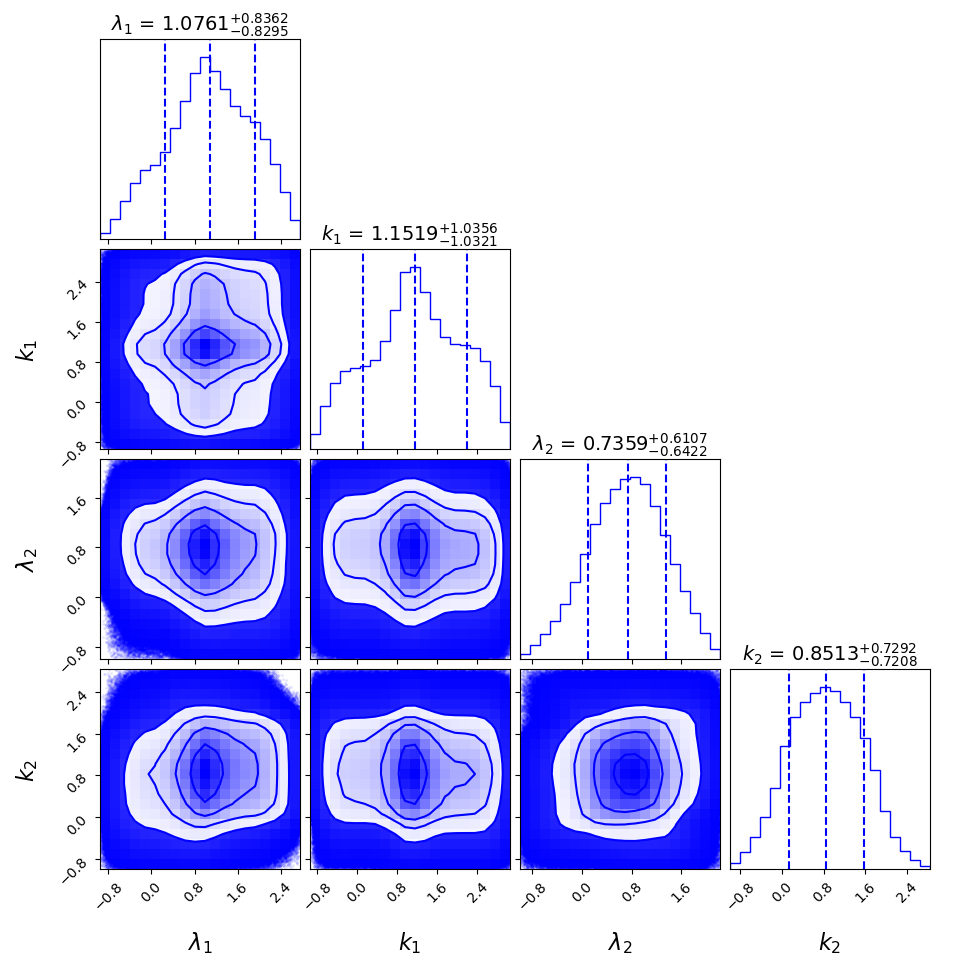}}
    \subfloat[$\log(\frac{M}{M_0})$ vs redshift plot] {\includegraphics[height=1.7in,width=1.7in]{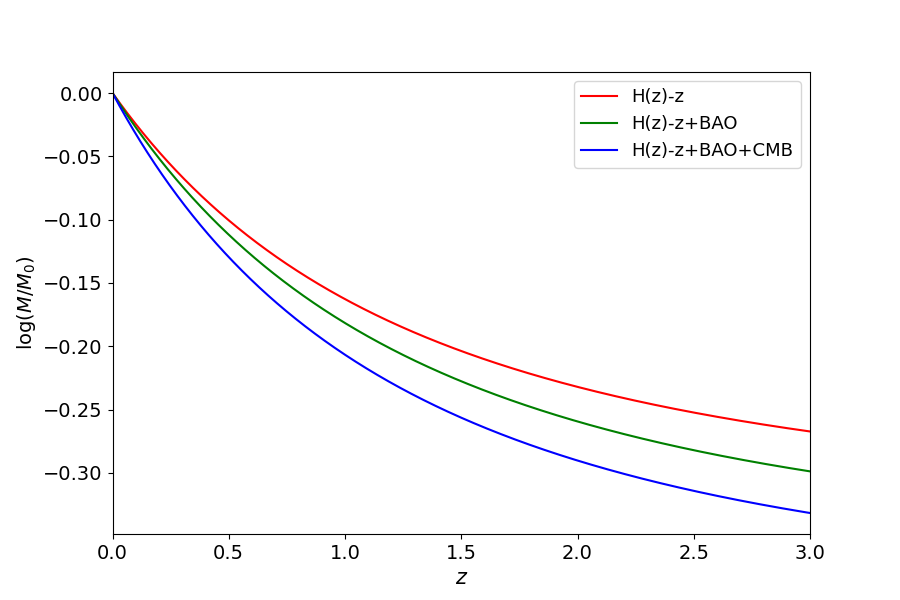}}
    \caption{\small{Fig. 1(a)-1(c) represent confidence contours in $\lambda_1-k_1~,~\lambda_1-\lambda_2~,~\lambda_1-k_2~, ~k_1-\lambda_2,~k_2-\lambda_2~\text{and}~k_1-k_2$ planes and individual distributions of free parameters $\lambda_1~,~k_1~, ~\lambda_2~\text{and}~k_2$. 1(d) represents the variation of the $\log\left\{\frac{M}{M_0}\right\}$ due to the accretion of BRB type DE onto Schwarzschild BH.}}
\end{figure}

In figure 1a to 1c, we plot confidence contours of all possible pairs chosen from the four free parameters $\lambda_1~,~k_1~, ~\lambda_2~\text{and}~k_2$. To constrain them, we have taken the data from table 1 of \cite{Biswas:2025bve} and the references therein. Best fits of these parameters are noted in table 1. Fig 1a is drawn for simple OHD. Fig 1b and 1c are drawn respectively for OHD+BAO and OHD+BAO+CMB constraints. In fig 1a, when the posterior distribution of a parameter ($\lambda_1$) has a narrow central peak but a wide full width at half maximum (FWHM), it reflects a very specific kind of inference. A narrow peak signifies that the posterior has a sharply defined most probable value (mode), i.e.,  the data highly favors one configuration. A broad half maximum width points that the likelihood remains relatively flat around it, meaning that a wide range of nearby values are almost equally acceptable to the data, although the mode is clear. This typically happens when the parameter enters the cosmological model nonlinearly or when it is degenerate with other parameters. So, the posterior density looks like a tall, thin peak, sitting atop a broad plateau, strongly peaked locally, but with weak curvature far from the maximum. 

In fig 1a, a narrow posterior peak accompanied by a wide FWHM indicates that the universe dynamically favors a specific parameter value, yet observational data do not sharply exclude nearby configurations. This implies that the physical process is robustly centered but weakly discriminative, such that the cosmic expansion behaves almost identically for a range of parameter values surrounding the best fit one. In the BRB model, the same expansion history $H(z)$ can often be reproduced by small compensating shifts among parameters, for example between $\lambda_1$, $k_1$, and $\Omega_m$. Consequently, the model is locally constrained (the peak is sharp) but globally degenerate (the plateau around the peak is broad). In terms of the Fisher information, this situation corresponds to $\frac{\partial^2 \ln \mathcal{L}}{\partial \omega_1^2}\bigg|_{\text{mode}}$ being large, but higher-order derivatives remaining small. Thus, the likelihood surface is steep near the maximum but decays slowly away from it, leading to a narrow central peak and a wide half maximum region.

For $k_1$, the posterior having a narrow peak on a wide plateau but being slightly left skewed implies that mode $>$ median $>$ mean, and the distribution resembles a tall, sharp spike superimposed on a broad, almost flat base with a longer left tail.

This indicates that highly localized best fit (narrow peak), i.e., the data prefer a very specific parameter configuration corresponding to the most probable cosmological dynamics. A wide neighborhood of the parameter space yields nearly equivalent expansion histories, reflecting degeneracy among model parameters. The mild tail toward smaller parameter values implies that weaker coupling or lower DE evolution remains statistically permissible, though less probable than the best fit configuration.

The broad base of the posterior implies that the expansion history, $H(z)$, and the luminosity distance, $D_L(z)$, remain almost unchanged even when the model parameters deviate from their best-fit values. This plateau corresponds to a degenerate manifold in the BRB parameter space, where variations in one parameter can be compensated by correlated shifts in another. Such compensation renders the cosmic acceleration structurally robust, as multiple parameter combinations yield macroscopically equivalent cosmological evolution.

The slight left skewness of the distribution signifies that smaller parameter values (representing weaker geometric coupling or smaller deviation from the $\Lambda$CDM limit) remain statistically permissible, although they are less probable than the best fit configuration. Physically, this asymmetry emerges because negative or small values tend to lead to phantom like or unstable cosmological branches that are weakly supported by observations, whereas moderate
positive values stabilize the late time attractor and provide a better fit to the data. Hence, the skewness reflects a thermodynamic preference for small but nonzero geometric deformation within the BRB framework.

$\lambda_2$ has a left skewed distribution. A left-skewed posterior distribution, characterized by a longer tail toward lower parameter values and a peak shifted toward higher values, implies that the cosmological data strongly favors smaller deviations from the best fit parameter while penalizing excessively large values.
Mathematically, such a distribution satisfies mean $<$ median $<$ mode.

In the context of the BRB DE model, this morphology indicates that the universe prefers a moderately deformed geometric or thermodynamic configuration, while smaller parameter values (e.g., weaker coupling or less non-extensivity) remain statistically acceptable but physically less favored.

Physically, the left skewness arises when the underlying dynamics contain a saturation effect, i.e., beyond a certain threshold, further reduction in the coupling parameter or deformation strength produces minimal change in the cosmological observables such as $H(z)$ or $D_L(z)$. As a result, the likelihood surface steeply declines for larger parameter values but decays more gradually for smaller ones. This asymmetry reveals that the BRB framework allows a range of
quasi-$\Lambda$CDM behaviors but exhibits an attractor like stabilization toward a mildly modified geometry, reflecting the dominance of nonlinear curvature or non-metricity corrections in the effective cosmic dynamics.

$k_2$'s symmetric distribution can be analysed physically, such a symmetric distribution reflects a dynamical balance between the competing effects of the geometric correction and the DE equation of state in the BRB framework. The parameter under consideration contributes to the modified expansion dynamics in a nearly linear or weakly nonlinear fashion, producing no preferential bias toward either stronger or weaker deviations from the $\Lambda$CDM limit.

This symmetry therefore signifies that the cosmic acceleration in the BRB model remains statistically neutral to small perturbations of that parameter, implying that the corresponding physical process operates close to a dynamical equilibrium. In other words, the data support an effectively stable configuration where the entropy production and geometric deformation balance each other, consistent with a quasi-extensive thermodynamic regime of late time cosmology.

In fig 1b and 1c, we have followed the distributions to be more or less similar to those of 1a. 

\begin{table}[ht]
\centering
\caption{\small{Best fit values of BRB model parameters from different datasets}}
\begin{tabular}{|>{\centering\arraybackslash}m{0.95cm}|
                >{\centering\arraybackslash}m{2cm}|
                >{\centering\arraybackslash}m{2cm}|
                >{\centering\arraybackslash}m{2.1cm}|}
\hline
\textbf{Tools} & $H(z)-z$ & $H(z)-z$ + $BAO$ & $H(z)-z$ + $BAO$+$CMB$ \\
\hline
\boldmath$\chi^2$ &  4655.117874  & 5416.228601 & 14610.907613 \\
\hline
\boldmath$\lambda_1$ & $0.9378^{+0.8667}_{-0.8638}$ & $0.9239^{+0.8649}_{-0.8678}$ & $1.0761^{+0.8362}_{-0.8295}$\\
\hline
\boldmath$\lambda_2$ & $0.9978^{+0.5274}_{-0.5604}$ & $1.0093^{+0.8649}_{-0.8678}$ & $0.7359^{+0.6107}_{-0.6422}$ \\
\hline
\boldmath$k_1$ & $1.0670^{+1.0114}_{-1.0206}$ & $1.1089^{+1.0358}_{-1.0360}$ & $1.1519^{+1.0356}_{-1.0321}$ \\
\hline
\boldmath$k_2$ &  $0.8287^{+0.7565}_{-0.7446}$ &  $0.8299^{+0.7577}_{-0.7446}$  &  $0.8513^{+0.7292}_{-0.7208}$\\
\hline
\end{tabular}
\end{table}

For DE accretion, we will consider a perfect fluid following the EoS $p=p(\rho)$, and the energy momentum tensor of the DE will be,
\begin{equation}\label{stree energy tensor}
    T_{mn}=(p+\rho)u_mu_n-pg_{mn}~~~,
\end{equation}
where $\rho$ and p are the DE density and pressure, respectively.

Now, $u^m\equiv \frac{dx^m}{d\tau_m}$ is the four speed component for the $m^{th}$ coordinate, $x^m$. $\tau$ is the proper time. Here, we assume $u_m u^n=1$ and that the spherical symmetry of the BH remains unaffected by the accreting DE. Without loss of generality, we take $u^1=u<0$ corresponding to a radial infall, and set $u^2=u^3=0$ due to symmetry considerations. The derivation of the velocity components then follows as 
\begin{align}
    u_1=\frac{u}{\mathcal{A}(r)}~,~u^0=\frac{\sqrt{{\mathcal{A}}(r)+u^2}}{{\mathcal{A}}(r)}~\text{and}~ u_0=-\sqrt{{\mathcal{A}}(r)+u^2}~,
\end{align}
where $\mathcal{A}(r)$ denotes the lapse function of Schwarzschild BH.
\begin{equation}
    \mathcal{A}(r) = 1- \frac{2 M}{r}~~~,
\end{equation}
with $M$ being the mass of the central gravitating Schwarzschild complex object.

In formulating the continuity equation, we take into account the corresponding components of the energy momentum tensor $T^{n}_{m}$ as
 \begin{equation}
 \left.{
 \begin{split}
    T^0_0 &=(\rho + p)u^0 u_0 + p =-\rho - \frac{u^2}{{\mathcal{A}}(r)} (\rho + p)~,\\
    T^1_1 &= (\rho + p) u^1 u_1 + p =(\rho + p)\frac{u^2}{{\mathcal{A}}(r)} + p~,\\
    T^2_2 &=T^3_3=p  ~\text{and}~T_0^1 = -u(\rho + p) \sqrt{{\mathcal{A}}(r) +u^2} ~~~.
 \end{split}
 }\right\rbrace
 \end{equation}
 
We will now involve the explicit BRB DE model effects through the conservation of mass flux $\frac{\partial J^m}{\partial x^m} = 0$ which can be represented in terms of the four-current that can be written as
\begin{align}
    J^m = \rho \begin{pmatrix} c \\ \dot{x}^i \end{pmatrix} = \begin{pmatrix} c \\ u \\ 0 \\0 \end{pmatrix} 
    \implies \rho~ u~ r^2 = \zeta_0^{BRB} \text{ (a constant)}
\end{align}
and choosing the $0^{th}$ component of $T^{n}_{m ; n} = 0$, (; signifies covariant differentiation) i.e., $T_{t;\nu}^\nu =0$, and we obtain the following
\begin{equation}\label{components of T mu nu}
\left.{
\begin{split}
    0 =& T^{n}_{0;n} = T^0_{0 ; 0} + T^1_{0 ; 1} + T^2_{0 ; 2} + T^3_{0 ; 3} \\
    \implies &u r^2 (\rho + p) \sqrt{{\mathcal{A}}(r)+u^2}M^{-2}=\zeta_1^{BRB}~~~,
\end{split}
}\right\rbrace
\end{equation}
where $\zeta_1$ is an arbitrary constant of integration.

Another repeatedly addressed EoS like CPL implies nonvanishing DE accretion at high redshift, artificial enhancement or suppression of BH mass in the early universe. This is unphysical because DE should not significantly affect BH growth during radiation/matter domination. For JBP accretion, early time effects are suppressed. This enforces a fixed redshift profile, underestimates late time dynamical effects and reduces sensitivity in the accretion integral. Thus it can miss physically relevant late-time mass growth. For BRB, $1+w(z)\rightarrow 0$ as $z\rightarrow \infty$. This implies $\dot{M}\rightarrow 0~~(z\gg 1)$. Hence BH growth is driven only by matter/radiation at early times. DE accretion becomes relevant only in the late universe. This matches physical expectations.

In the context of energy momentum conservation, we consider $u^m T^n_{m; n} =0$, corresponding to the projection of the conservation law along the fluid’s four-velocity. This yields the following relation \cite{debnath2015accretion}
\begin{equation}\label{fluid four velocity}
    ur^2 M^{-2} \exp \left\{ \int^\rho_{\rho_\infty} \frac{d {\rho}'}{{\rho}' + p({\rho}')}\right\} = - \zeta_2^{BRB}~~~.
\end{equation}
$\zeta_2^{BRB} > 0$ represents also an integrating constant, related to the energy flux \cite{debnath2015accretion, babichev2004black, babichev2005accretion,babichev2013black}. $\rho$ and $\rho_{\infty}$ help to understand the nature of the density, we examine its behavior both at finite distances and in the asymptotic limit as it approaches infinity. For this purpose, we introduce a dimensionless radial distance parameter. $x=\frac{r}{M}$ the eq. \eqref{fluid four velocity} turns into 
\begin{equation}\label{lambda 2}
   ux^2 \exp \left\{ \int^\rho_{\rho_\infty} \frac{d {\rho}'}{{\rho}' + p({\rho}')}\right\} = - \zeta_2^{BRB}~~~.
\end{equation}
Calculating the value of $ux^2$\cite{Biswas:2025aku} we will get the following form of $\zeta_2^{BRB}$
 \begin{equation}
     \zeta_2^{BRB} = \frac{x_c^3{\mathcal{A}}'(x_c)}{2} \exp \left\{ \int^{\rho_c}_{\rho_\infty} \frac{d {\rho}'}{{\rho}' + p({\rho}')}\right\}~~~,
 \end{equation}
where $' \equiv \frac{d}{dx}$ denotes the differentiation with respect to $x$. The critical point is $x_c$. Hence $\mathcal{A}^\prime (x_c) = + \frac{2}{x_c^2}$. So, for different redshift parameterizations, we can obtain the constants $\zeta_2^{BRB}$.

Moreover, the rate of change of the mass of BH is to be obtained as

\begin{equation}
     \dot{M}=-4 \pi \zeta_1^{BRB} M^2 = 4 \pi \zeta_2^{BRB} M^2 \left\lbrace{\rho_\infty + p\rho_\infty)}\right\rbrace~~~.
\end{equation}

Several articles \cite{Abbas_2013, Babichev_2008, Jamil_2010, Jamil_2010_GRG, Dutta_2019} suggest that the rate of change of mass can be calculated, but does not satisfy the dominant energy condition whenever we can apply the EoS $p=\omega\rho$.

Hence, \begin{equation}
    \dot{M} = 4 \pi \zeta_2 M^2 [\rho+p(\rho)]~~~.
\end{equation}

Sign of $\dot{M}$ is determined by $(\rho+p)$. For quintessence $-1<\frac{p}{\rho}<-\frac{1}{3}$, and hence $\dot{M}>0$. Once the system crosses the phantom barrier, where the EoS satisfies $\left({\frac{p}{\rho}=-1}\right)$, the inequality $p+\rho<0$ holds, leading to $\dot{M}<0$.

\begin{equation}
   \dfrac{d M}{d \rho} =-\frac{4 \pi \zeta_2^{BRB} M^2}{3H^2} \Rightarrow M =\frac{M_0}{1+ \frac{4 \pi \zeta_2^{BRB} M_0}{3} \int\limits_{\rho}^{\rho_0} \frac{d \rho }{H}}~~.
\end{equation}
The constant $\zeta_2^{BRB}$ is calculated using the equation \eqref{Omega} to derive the expression and we solve this numerically as the function of density over a large range onto the solution of Schwarzschild BH.
\begin{equation}
    \frac{d\rho^\prime}{\rho^\prime+p(\rho^\prime)}=\frac{d\rho^\prime}{\left\lbrace{\frac{\lambda_1}{1+k_1+z}+\frac{\lambda_2 z}{\left(1+k_2+z\right)^2}}\right\rbrace\rho^\prime}~~~.
\end{equation}

Now, for our model, the equation \eqref{E_square equation} takes the following form
\begin{equation} \label{E_square equation}
\begin{split}
    E^2(z) \equiv \frac{H^2 (z) }{H_0^2} = \Omega_{b0} (1+z)^3 +\Omega_{DE0} \beta\frac{(1+ak_2)^{\frac{3\lambda_2}{k_2^2}}}{(1+ak_1)^{\frac{3\lambda_1}{k_1}}} \\
    \times \exp{\left\lbrace{{\frac{3\lambda_2(1+k_2)}{k_2^2(1+ak_2)}}}\right\rbrace} + \Omega_{r0} (1+z)^4+ \Omega_{k 0}(1+z)^2~~~,
\end{split}
\end{equation}
where, $\beta=\frac{(1+k_1)^{\frac{3\lambda_1}{k_1}}}{(1+k_2)^{\frac{3\lambda_2}{k_2^2}}} \exp{\left(-\frac{3\lambda_2}{k_2^2}\right)},~~\Omega_{DE0}=1-\Omega_{b0}-\Omega_{r0}$. Here $M_0 \left(M(z=0)\right)$ is the present day mass of the BH.

Hence we obtain 
\begin{equation}
\begin{split}
    M &= M_0 \Bigm/ \left[{1+ \frac{4 \pi \zeta_2^{BRB} M_0}{3}}\right. \\
    &\left.{\times \int\limits_{\rho_0}^\rho d \left({\rho_{\phi0}\frac{(1+k_1)^{{3\lambda_1}{k_1}}}{(1+k_2)^{\frac{3\lambda_2}{k_2^2}}}\exp{\left\lbrace{-\frac{3\lambda_2}{k_2^2}}\right\rbrace} }\right.}\right. \\
    &\left.{ \left.{\times \frac{(1+ak_2)^{\frac{3 \lambda_2}{k_2^2}}}{(1+ak_1)^{\frac{3\lambda_1}{k_1}}}\exp{\left\lbrace{{\frac{3\lambda_2(1+k_2)}{k_2^2(1+ak_2)}}}\right\rbrace}}\right) \Bigm/ }\right.\\
    &\left.{H_0\left\lbrace{ \Omega_{rad0}a^{-4}+\Omega_{dm0}a^{-3}+\Omega_{\phi0}\beta \frac{(1+ak_2)^{\frac{3\lambda_2}{k_2^2}}}{(1+ak_1)^{\frac{3\lambda_1}{k_1}}}}\right.}\right.\\
    &\left.{\left.{\times \exp{\left\lbrace{{\frac{3\lambda_2(1+k_2)}{k_2^2(1+ak_2)}}}\right\rbrace} +\Omega_{k 0}(1+z)^2}\right\rbrace^{\frac{1}{2}}}\right]~~~.
\end{split}
\end{equation}

In Fig.1(d), we have plotted $\log_{10}\left( \frac{M^{BRB}}{M_0^{BRB}}\right)$ vs. $z$ for growing mass. We follow the mass to increase in the past. Mass of the BH is found to gain almost 55\% of its present time mass in the era from redshift 3 to redshift zero. The observed increase of $\log_{10}[M(z)/M_0]$ since $z=3$ reflects the cumulative growth of supermassive BHs through both gas accretion and hierarchical mergers. Accretion, whether radiatively efficient or inefficient, can produce exponential-like mass growth when sustained, while mergers contribute discrete mass increments; the net observed growth is the sum of these channels. The effective growth rate depends on the radiative efficiency $\epsilon$, the Eddington ratio and duty cycle combined as $\lambda_{\rm eff}$, and the cosmic timing, with the most rapid accretion typically occurring near $z\sim2$--$3$ and slower, merger dominated growth toward $z=0$. Observational selection effects and systematics can bias the measured $\log(M/M_0)$ evolution, so the quantitative interpretation requires careful consideration of these factors. Overall, the increasing trend is consistent with standard hierarchical growth and quasar driven supermassive BH(SMBH) assembly, though the precise growth depends on the adopted accretion and merger parameters.

The mass growth equation integrates as
$$M^{-1}(z)-M^{-1}(z_i)\propto \int z_{z_i}\frac{\rho_{DE}(z')\left[1+w(z')\right]}{H(z')\left(1+z'\right)}dz'$$
CPL can make this integral diverge or saturate incorrectly. JBP often suppressed it too smoothly. BRB keeps the integrand finite, smooth and sharply peaked at late times. Hence the mass ratio $\frac{M_0}{M_i}$ is more stable and physically meaningful. Observationally, most SMBH mass growth is followed to occur at $z\lesssim 2$. Earlier epochs hardly show any influence of DE onto BH accretion. BRB model automatically enforces this behavior and also avoids artificial early enhancement seen in CPL based accretion studies. Thus better consistency with astrophysical details is expected.


{\bf {Conclusion :}} DE accretion is a physically meaningful problem because the DE fluid, described by its stress-energy tensor $T_{\mu\nu}$, can exchange energy and momentum
with strong gravitational fields. Its inflow modifies the black hole mass evolution as $\dot{M} = 4\pi r^{2} T^{r}{}_{t}$, revealing whether the horizon grows for quintessence-like DE ($w > -1$) or decreases for phantom fields ($w < -1$). This process thus serves as a probe of the microscopic nature of DE, a test of energy-momentum conservation in curved spacetime, and a means to examine the validity of black hole thermodynamics in the presence of cosmic fluids.

The BRB type DE model proposes a generalized
redshift-dependent equation of state, $\omega(z) = \omega_0 + \omega_1 f(z)$, which extends beyond the constant-$\omega$ and CPL forms by introducing a smooth functional evolution with redshift. The model successfully mimics the $\Lambda$CDM behaviour at the present epoch ($\omega \simeq -1$) while allowing small deviations characterized by $\omega_1 \neq 0$. When constrained by observational $H(z)$ data, the parameters indicate that DE remains close to a cosmological constant but may evolve slightly with time, potentially leading to a future deceleration phase. Thus, the BRB model provides a phenomenological bridge between constant-$\omega$ and fully dynamical DE scenarios, offering a flexible yet observationally consistent description of late-time cosmic acceleration.

Two free parameters $\lambda_1$ and $k_1$ are followed to have distributions with narrow peak sitting on a wide plataeu. The universe exhibits a preferred DE evolution, corresponding to a definite dynamical attractor. However, this attractor is surrounded by a plateau of quasi-degenerate states that yield nearly identical cosmic expansion histories. This structure reflects the insensitivity of current cosmological observations
to subtle dynamical variations in the DE sector, a wellknown limitation in reconstructing fine features of the late-time acceleration. 

$${
\begin{split}
    M(t) &= M_{\rm seed}\times \exp\Bigg[\frac{1-\epsilon}{\epsilon}\,\lambda_{\rm Edd}\,\frac{t}{t_{\rm S}} \Bigg] ~\text{and}\\
    t_{\rm S} &\simeq 4.5\times10^{7}\,\mathrm{yr}\left(\frac{\epsilon}{0.1}\right),
\end{split}}
$$

where $\epsilon$ is the radiative efficiency, $\lambda_{\rm Edd}\equiv L/L_{\rm Edd}$ is
the Eddington ratio (often replaced by an effective $\lambda_{\rm eff}=\lambda_{\rm Edd}\times f_{\rm duty}$ to include duty cycle)
and $t_{\rm S}$ is the Salpeter time.  

$$
\log_{10}\frac{M(z)}{M_0}
= \frac{1}{\ln 10}\,\frac{1-\epsilon}{\epsilon}\,
\lambda_{\rm eff}\,\frac{\Delta t(z\to0)}{t_{\rm S}}~~~,
$$
with \(\Delta t(z\to 0)\) the cosmic time elapsed from redshift \(z\) to today.

\bibliographystyle{ieeetr} 

\bibliography{references}

\end{document}